\documentclass[preprint,showpacs,preprintnumbers,amsmath,amssymb]{revtex4}
\usepackage{graphicx,dcolumn,bm}

\begin{document}


\title{Crossover from classical to quantum behavior of Duffing oscillator \\ through ``pseudo-Lyapunov exponent''}

\author{Yukihiro Ota}
 \email{ota@suou.waseda.jp}
\affiliation{
Department of Physics, Waseda University, Tokyo 169--8555, Japan }
\author{Ichiro Ohba}
 \email{ohba@waseda.jp}
\affiliation{%
Department of Physics, Waseda University, Tokyo 169--8555, Japan \\ Kagami Memorial Laboratory for Material Science and Technology, Waseda University, Tokyo 169--0051, Japan \\ Advanced Research Center for Science and Technology, Waseda University, Tokyo 169--8555, Japan
}%

\date{\today}

\begin{abstract}
We discuss the quantum--classical correspondence in a specific dissipative chaotic system, Duffing oscillator.
We quantize it on the basis of quantum state diffusion (QSD) which is a
 certain formulation for open quantum systems and an effective tool for
 analyzing complex problems numerically.  
We consider a sensitivity to initial conditions, `` pseudo-Lyapunov
 exponent '', and investigate it in detail,  varying Planck constant effectively.
We show that in a dissipative system there exists a certain critical stage in which the crossover from classical to quantum behavior occurs. 
Furthermore, we show that an effect of dissipation suppresses the
 occurrence of chaos in the quantum region, while it, combined with the
 periodic external force, plays a crucial role in the chaotic behaviors of classical system. 
\end{abstract}

\maketitle

\section{\label{Sec:Intro}Introduction}
The quantum-classical correspondence is a very important problem related to the foundation of quantum mechanics. 
However, it is difficult to consider this correspondence if a classical system which we should quantize is a chaotic system (or generic nonintegrable system). 
For, the phenomena in the quantum mechanics corresponding to the chaos in the classical mechanics can not exist generally due to quantum mechanical dynamics\cite{Nakamura, Peres}.

Nevertheless, the various works have been done in Hamiltonian systems.
The central task for studying the quantum-classical correspondence in Hamiltonian chaotic systems is how the measured quantities in a quantum system relate to the information of trajectories in its corresponding classical system. 
The study in this point is very fruitful\cite{Casati, Haake, Gutzwiller}. 

However, researches limited in Hamiltonian systems are not sufficient to study the quantum--classical correspondence in chaotic systems.
There exist another types of chaos not having Hamiltonian. 
This phenomenon can occur in dissipative systems that a definite Hamiltonian does not exist.
In this paper, we discuss the quantum--classical correspondence for Duffing oscillator in the view of an open quantum system.

There are several reasons why we study this system in order to consider the dissipative quantum chaos. 
The authors in Ref.~\cite{Brun} reported interesting results for the same system as our model. 
Especially, the proof of existence of chaotic behavior in their classical limits is very important. 
We think, however, their results are insufficient to discuss the crossover from classical to quantum behavior.
In fact, we show that a new method, an analysis based on ``pseudo--Lyapunov exponent'', is possible to clarify the crossover behaviors.
Moreover, there are experimental propositions for dissipative quantum anharmonic oscillator which are not identical with the Duffing oscillator. 
In Ref.~\cite{Adamyan}, the analysis is worked in such a model.
Furthermore, the investigation of this problem is related to not only the quantum chaos but also the several fundamental problems in the quantum mechanics, for example, the influence of dissipation on quantum tunneling phenomena\cite{Caldeira} and the quantum stochastic resonance\cite{Adamyan2}.
Finally, it also has interesting features in the classical mechanics\cite{Kim}.
The other system for the dissipative quantum chaos, e.g., the dissipative kicked top, is studied in Ref.~\cite{Braun, Dittrich}. 

The main aim is what happens in the stage between quantum and classical
regions in a dissipative system, as the Planck constant changes effectively.  
It is expected that the various phenomena related to quantum--classical
correspondence should occur.  
But it is not clear what kind of quantity we should investigate to find such a crossover.
We examine a quantity sensitive to the initial condition and define ``pseudo--Lyapunov exponent'' as its candidate. 
Then we will discuss such a crossover in the quantized Duffing oscillator. 
The effective Planck constant $\beta$ and the effective Planck cell play an important role in this analysis. 

We obtain several fruitful results.
First, we find that ``pseudo--Lyapunov exponent'' is certainly positive when the system of Duffing oscillator is in a classical region. 
Furthermore, we show that there is a certain clear critical stage in which the crossover from classical to quantum behavior occurs.
We find also that the effect of dissipation is different between classical and quantum regions; this, together with a periodic external force, plays the essential role for chaotic behaviors in the classical region but suppresses the occurrence of chaos in a quantum region.

The contents of this paper are as follows. 
In Sec.~\ref{Sec:Model}, we explain the model, Duffing oscillator. 
Especially, we explain the phenomenological description of open quantum systems, quantum state diffusion (QSD). 
Our method is identical with that in Ref.~\cite{Brun}.
We explain in detail the introduction of scaling parameter $\beta$ which is very important for the investigation of crossover behavior.
In Sec.~\ref{Sec:CR}, we show several numerical results without averaging over the ensemble for complex Wiener process used in QSD.
These methods give the proper results around $\beta=0.01$.
Such a reserch has been already studied in Ref.~\cite{Brun}.
The results are not first realizations, but are important to explain our motivation of the analysis in the next section. 
In Sec.~\ref{Sec:Cross}, we show the main results in this paper. 
We introduce a quantity sensitive to initial conditions, and find that there exists a clear crossover from classical to quantum behavior as $\beta \to 1$.
In Sec.~\ref{sec:Summ}, we summarize this paper. 
In Appendix \ref{app:A}, we derive an equation used in Sec.~\ref{subsec:QRDQR}.

\section{\label{Sec:Model}Model}
In this section, we first review the classical Duffing oscillator briefly.
We quantize the Duffing oscillator as an open quantum system phenomonologically, using QSD. 
This method is identical with that in Ref.~\cite{Brun}.

The equation of motion for classical Duffing oscillator is the following: 
\begin{eqnarray}
\ddot{x} + 2\Gamma \dot{x} + x^{3} - x = g \cos(\Omega t).
\label{Eq:cDuff}
\end{eqnarray}
It is known that the various behaviors can occur depending on the set of parameters $\Gamma$, $g$ and $\Omega$.
The chaotic behavior appears in the case of $\Gamma = 0.125$, $g=0.3$ and $\Omega=1.00$\cite{Guckenheiner}. 
We find a strange attractor in Poincar\'e surface, which is obtained by putting $(x,\,p)$ in a phase space by every interval of $2\pi /\Omega$. 
Hereafter we use this set of parameters.
The appearance of strange attractor in such a surface is one of the properties in dissipative chaotic systems. 

We describe phenomenologically the dynamics of such a system without a well--defined Hamiltonian; we regard it as an open quantum system. 
We assume the Markovian dynamics and choose the effect of dissipation phenomenologically 
\footnote{
The validity of Markovian approximation for nonlinear oscillator is discussed in Ref.~\cite{Alicki}, using the specific microscopic model.
}.
Then the dynamics of system is described by QSD \cite{Gisin,Gisin1,Gisin2,Gisin3,Percival,Brun,Schack} in which the pure state vector of system evolves according to the It\^o stochastic differential equation: 
\begin{eqnarray}
\rvert d\psi \rangle &=& -\frac{i}{\hbar} \hat{H} \rvert \psi \rangle dt + \bigg( \langle \hat{L}^{\dagger} \rangle \hat{L}-\frac{1}{2} \hat{L}^{\dagger} \hat{L} \nonumber \\ & & - \frac{1}{2} \langle \hat{L}^{\dagger} \rangle \langle \hat{L} \rangle \bigg) \rvert \psi \rangle dt +\bigg( \hat{L}-\langle \hat{L} \rangle \bigg) \rvert \psi \rangle d\xi, \label{Eq:qsd}
\end{eqnarray}
where
$\text{M}\{ d\xi \}=0$, $\text{M}\{d\xi d\xi \}=0$ and $\text{M}\{ d\xi^{\ast} d\xi \} =dt$. 
$d\xi$ describes the increment of complex Wiener process and $\text{M}$ expresses the ensemble average for it. 
$\langle \bullet \rangle$ represents $\langle \psi |\bullet |\psi \rangle$. 
The quantum expectation value for an operator $\hat{O}$ is represented by 
$$
\text{M} \{ \langle \psi \lvert \hat{O} \rvert \psi \rangle \} = \text{Tr} \{ \hat{O}\rho \}, 
$$ 
where $\rho$ is a reduced density matrix for the system 
\footnote{
Notice that $\text{Tr}(\hat{O}^{n}\rho) \neq \text{M} (\langle \hat{O}^{n} \rangle )$ ($n \ge 2$) if $\hat{O}$ depends on a specific realization of $\xi (t)$, say, $\hat{O}=\hat{Q}-\langle \hat{Q} \rangle $. 
}.
$\hat{H}$ is a certain self-adjoint operator and it is called by {\it Hamiltonian} since the term related to $\hat{H}$ describes the unitary evolution. 
$\hat{L}$ is called by a Lindblad operator and describes the effect of dissipation.
The QSD is equivalent to the Lindblad master equation
\cite{Lindblad,Spohn}:
\begin{eqnarray}
\dot{\rho} = -\frac{i}{\hbar} [ \hat{H},\rho ] + \hat{L} \rho \hat{L}^{\dagger} - \frac{1}{2} \hat{L}^{\dagger} \hat{L} \rho - \frac{1}{2} \rho \hat{L}^{\dagger} \hat{L}. 
\label{Eq:Lindblad}
\end{eqnarray}
The Lindblad master equation is a quite general formulation for open quatum systems satisfying the Markovian dynamics, trace preserving and complete positivity.
The QSD is a very effective tool for numerical simulation of complex problems \cite{Brun}, compared with the description depending on the master equation. 
In this paper, we use the algorithm for QSD which the authors in Ref.~\cite{Schack2} invented. 
The QSD is also possible to explain a measurement processes in the quantum mechanics\cite{Ghiraradi, Goetsch}.

In order to describe the dynamics of Duffing oscillator in the quantum mechanics, we define the Hamiltonian $\hat{H}$ and a Lindblad operator $\hat{L}$ in the Eq.~(\ref{Eq:qsd}) as the followings:
\begin{subequations}
\label{Eq:qDuff}
\begin{eqnarray}
\hat{H}=\hat{H}_{D}+\hat{H}_{R}+\hat{H}_{ex},
\label{Eq:Hamiltonian}
\end{eqnarray}
\begin{eqnarray}
\hat{H}_{D} = \frac{1}{2m} \hat{p}^{2} + \frac{m \omega^{2}_{0}}{4l^{2}} \hat{x}^{4} - \frac{m \omega^{2}_{0}}{2} \hat{x}^{2},
\label{Eq:qDuffH}
\end{eqnarray}
\begin{eqnarray}
\hat{H}_{R}=\frac{\gamma}{2} (\hat{x} \hat{p} + \hat{p} \hat{x}), 
\label{Eq:reNH}
\end{eqnarray}
\begin{eqnarray}
\hat{H}_{ex}=- gml \omega_{0}^{2} \hat{x} \cos(\omega t),
\label{Eq:exH}
\end{eqnarray}
\begin{eqnarray}
\hat{L}=\sqrt{\frac{m\omega_{0} \gamma}{\hbar}}\; \hat{x} + i \sqrt{\frac{\gamma}{m\omega_{0} \hbar}}\; \hat{p}. 
\label{Eq:Lop}
\end{eqnarray}
\end{subequations}
The Eq.~(\ref{Eq:reNH}) means the strength renormalization for coupling of interaction between system and environment. 
In fact, if we implement the canonical transformation, $x \to x$ and $p \to p-\gamma mx$, we can obtain the following equations:
$$ 
[\hat{x} ,\hat{p} ] \to [\hat{x} ,\hat{p} ], 
$$
\begin{eqnarray*}
 \hat{H}_{D} &+& \hat{H}_{R} \to \\& &  \frac{1}{2m} \hat{p}^{2} + \frac{m^{2} \omega_{0}^{2}}{4l^{2}} \hat{x}^{4} - \frac{m^{2} \omega_{0}^{2}}{2} \left( 1 + \frac{\gamma^{2}}{\omega_{0}^{2}} \right) \hat{x}^{2},
\end{eqnarray*}
$$
\hat{L} \to \sqrt{\frac{m\omega_{0}\gamma}{\hbar}}
\left(1-i\frac{\gamma}{\omega_{0}} \right) \hat{x} + i
\sqrt{\frac{\gamma}{m\omega_{0}\hbar}} \hat{p}.
$$
The Eq.~(\ref{Eq:exH}) means the external force depending on time periodically. 
Notice that the right hand side of Eq.~(\ref{Eq:Lindblad}) is independent of the time. 
Therefore, it is a difficult problem how the generator depends on the time. 
We determine simply it so as reproduce the external force in the equation of expectation values.

We rewrite Eq.~(\ref{Eq:qsd}) into the dimensionless form:
\begin{eqnarray}
\rvert d\psi \rangle &=& -\frac{i}{\hbar} \hat{H}_{\beta} \rvert \psi \rangle d\tau + \bigg( \langle \hat{K}^{\dagger} \rangle \hat{K}-\frac{1}{2} \hat{K}^{\dagger} \hat{K} \nonumber \\ & & - \frac{1}{2} \langle \hat{K}^{\dagger} \rangle \langle \hat{K} \rangle \bigg) \rvert \psi \rangle d\tau +\bigg( \hat{K}-\langle \hat{K} \rangle \bigg) \rvert \psi \rangle d\zeta ,\label{Eq:scalqsd}
\end{eqnarray}
where $\text{M} \{ d\zeta \}=\text{M} \{ d\zeta d\zeta \}=0$, $\text{M} \{ d\zeta ^{\ast} d\zeta \} =d\tau$, $\hat{H}_{\beta} \equiv \hat{H}/\hbar \omega_{0}$, $\hat{K} \equiv \hat{L}/\sqrt{\omega_{0}}$, $\tau \equiv \omega_{0}t$ and $d\zeta \equiv \sqrt{\omega_{0}} d\xi$.
We define the unit of energy as $\hbar \omega_{0}$.
Moreover, we redefine $\hat{x}$ and $\hat{p}$ as $\hat{Q} \equiv \sqrt{m\omega_{0}/\hbar}\; \hat{x}$ and $\hat{P} \equiv \sqrt{1/m\omega_{0}\hbar}\; \hat{p}$, respectively.
Thus we obtain the dimensionless Hamltonian $\hat{H}_{\beta}$ and Lindblad operator $\hat{K}$:
\begin{subequations}
\label{Eq:scalqDuff}
\begin{eqnarray}
\hat{H}_{\beta} = \hat{H}_{D} + \hat{H}_{R} + \hat{H}_{ex},
\end{eqnarray}
\begin{eqnarray}
\hat{H}_{D} = \frac{1}{2} \hat{P}^{2}+\frac{\beta^{2}}{4} \hat{Q}^{4}-\frac{1}{2} \hat{Q}^{2},
\end{eqnarray}
\begin{eqnarray}
\hat{H}_{R} = \frac{\Gamma}{2} \left(\hat{Q} \hat{P} + \hat{P} \hat{Q} \right),
\end{eqnarray}
\begin{eqnarray}
\hat{H}_{ex}=-\frac{g}{\beta} \hat{Q} \cos(\Omega t),
\end{eqnarray}
\begin{eqnarray}
\hat{K} = \sqrt{\Gamma} \left( \hat{Q} + i\hat{P} \right),
\end{eqnarray}
\end{subequations}
where $\Omega \equiv \omega/\omega_{0}$, $\Gamma \equiv \gamma/\omega_{0}$.
The $\beta^{2}$ is the ratio of $\hbar$ to the characteristic action of system, $ml^{2}\omega_{0}$: 
\begin{eqnarray}
\beta^{2} = \frac{\hbar}{ml^{2}\omega_{0}}. \label{Eq:beta}
\end{eqnarray} 
We can effectively change $\hbar$, varying $\beta^{2}$ in the numerical computation. 
Notice that the varying of $\beta$ is just the scale transformation for system: $\Delta Q / \text{M} \{ \langle \hat{Q} \rangle \}$ and $\Delta P / \text{M} \{ \langle \hat{P} \rangle \} $ should vanish when $\beta$ goes to zero, where $\Delta \hat{Q}^{2} \equiv ( \hat{Q}-\langle \hat{Q} \rangle )^{2} $, $\Delta \hat{P}^{2} \equiv (\hat{P}-\langle \hat{P} \rangle )^{2}$, $\Delta Q \equiv \sqrt{ \text{M} \{ \langle \Delta \hat{Q}^{2}\rangle \} }$ and $\Delta P \equiv \sqrt{ \text{M} \{ \langle \Delta \hat{P}^{2}\rangle \} }$
\footnote{
$\Delta Q ^{2}$ is not a usual quantum variance, $\text{Tr} \{ \hat{Q}^{2}\rho \} - ( \text{Tr} \{ \hat{Q}\rho \} )^{2}$.
But, we expect that it is equal to the usual variance when $\langle \hat{Q}\rangle$ does not depend on a specific realization of $\zeta(t)$.
This is achieved in the classical case, $\beta \approx 0$.}.

We consider $\beta=1$ for a moment. 
Using It\^o calculus, we obtain the following equations \cite{Gardiner}:
\begin{subequations}
\label{Eq:expect}
\begin{eqnarray}
d\langle \hat{Q} \rangle &=& \langle \hat{P} \rangle d\tau + \sqrt{\Gamma}\Big[ \Big\{ \Big(V_{Q}-\frac{1}{2}\Big) \nonumber \\ & & +iV_{QP} \Big\} d\zeta + c.c.\Big], \label{ex_sub1}
\end{eqnarray}
\begin{eqnarray}
d\langle \hat{P} \rangle &=& (-2\Gamma \langle \hat{P} \rangle -\langle \hat{F} \rangle +g\cos{(\Omega \tau)})d\tau \nonumber \\ & & + \sqrt{\Gamma}\Big[ \Big\{ V_{QP}+i\Big(V_{P}-\frac{1}{2}\Big) \Big\} d\zeta + c.c.\Big], \label{ex_sub2}
\end{eqnarray}
\end{subequations}
where $V_{Q} \equiv \langle \Delta Q^{2} \rangle $, $V_{P} \equiv \langle \Delta P^{2} \rangle $, $2V_{QP} \equiv \langle \Delta \hat{Q} \Delta \hat{P} + \Delta \hat{P} \Delta \hat{Q} \rangle$ and $\hat{F} \! \equiv \! \hat{Q}^{3} - \hat{Q}$. 
Notice that, if we can approximately neglect moments more than second order, then we can see that Eq.~(\ref{Eq:expect}) reproduces the equation of motion for classical Duffing oscillator.
This is one of physically useful advances in the QSD. 
In such a case, a specific realization of stochastic process does not
allow any deviation from the classical behavior.
Therefore, we are able to guess reasonably that the classical regions
would be robust for a specific realization of stochastic process.

\section{\label{Sec:CR}behavior in the classical region}
The discussion in Sec.~\ref{Sec:Model} allows us to obtain the proper results around $\beta=0.01$ without averaging over the ensemble for complex Wiener process $\zeta(\tau)$.
See, Eq.~(\ref{Eq:expect}).
In this section, we show the several numerical results without averaging over $\zeta (t)$.
These results are useful to understand the behavior of quantized Duffing
oscillator intuitively for different values of $\beta$.

\begin{figure*}[htbp]
  \centering
  \begin{tabular}{cc}
   \includegraphics[scale=.5]{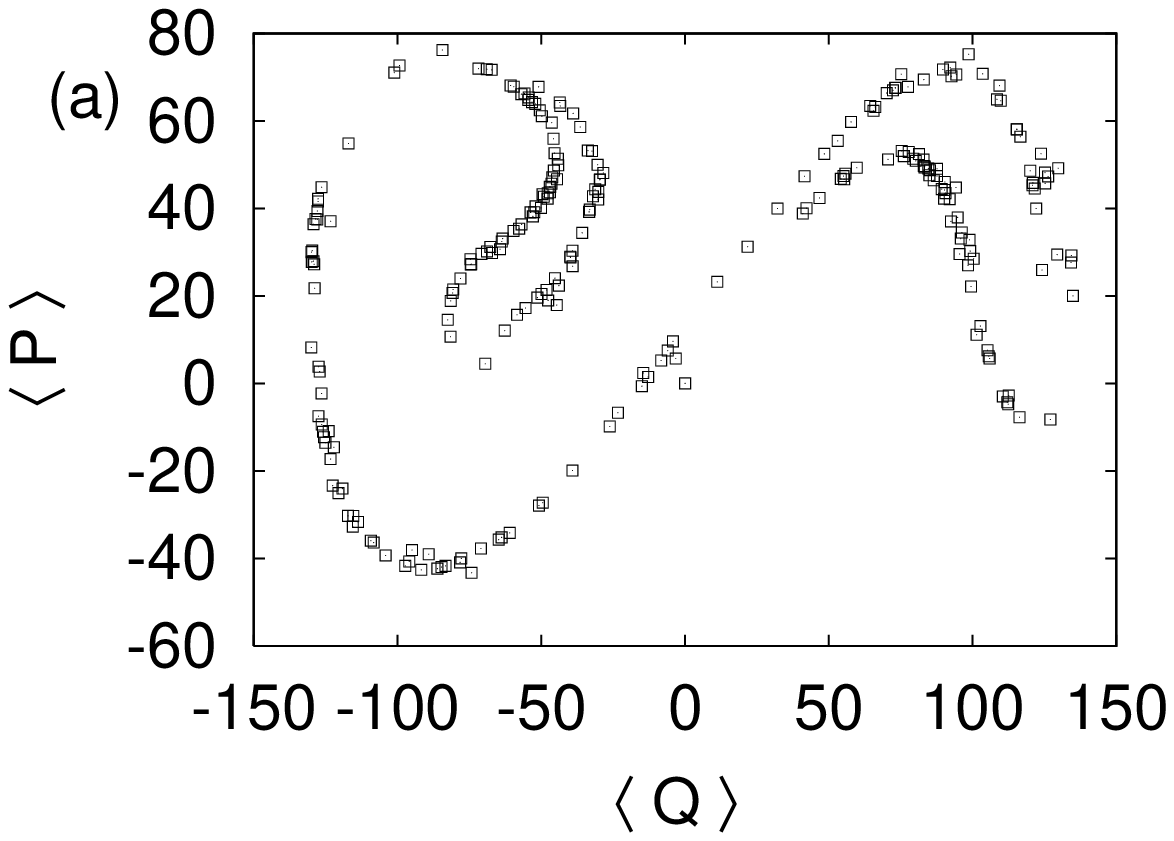} & 
   \includegraphics[scale=.5]{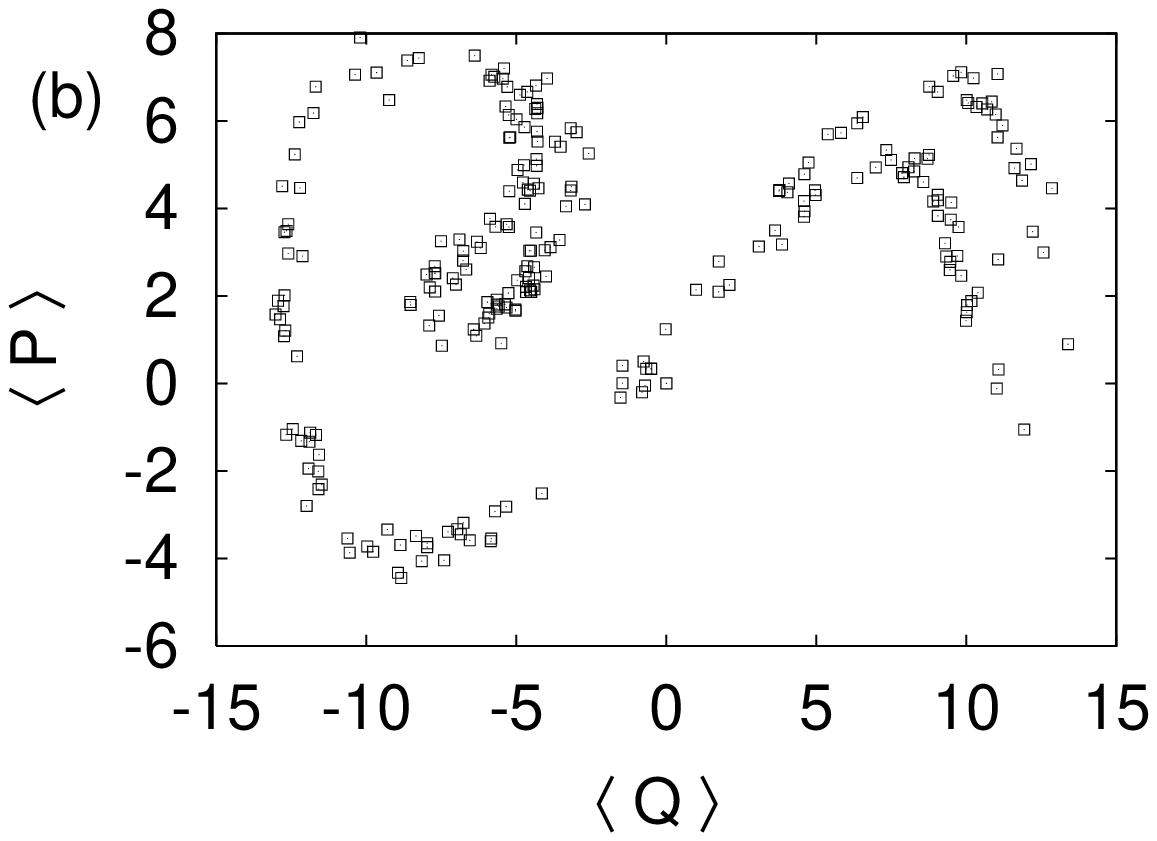}  \\ 
   \includegraphics[scale=.5]{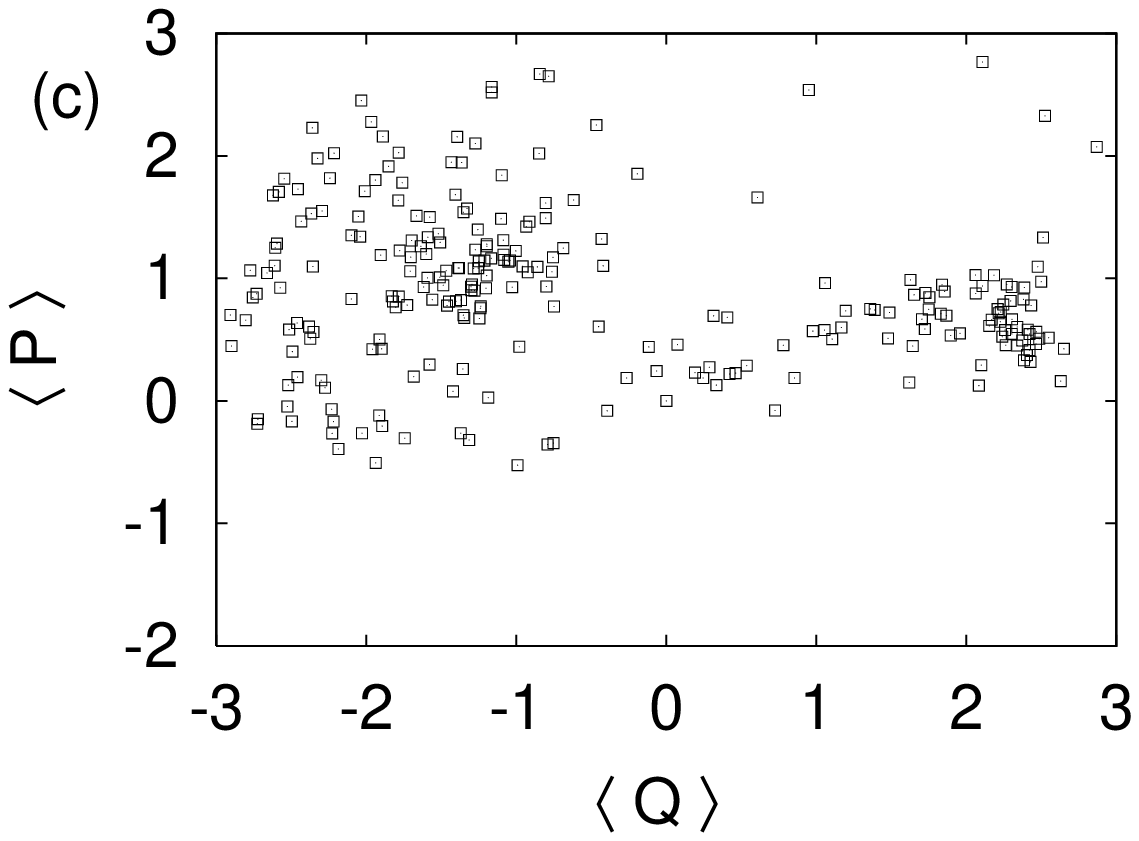}  & 
   \includegraphics[scale=.5]{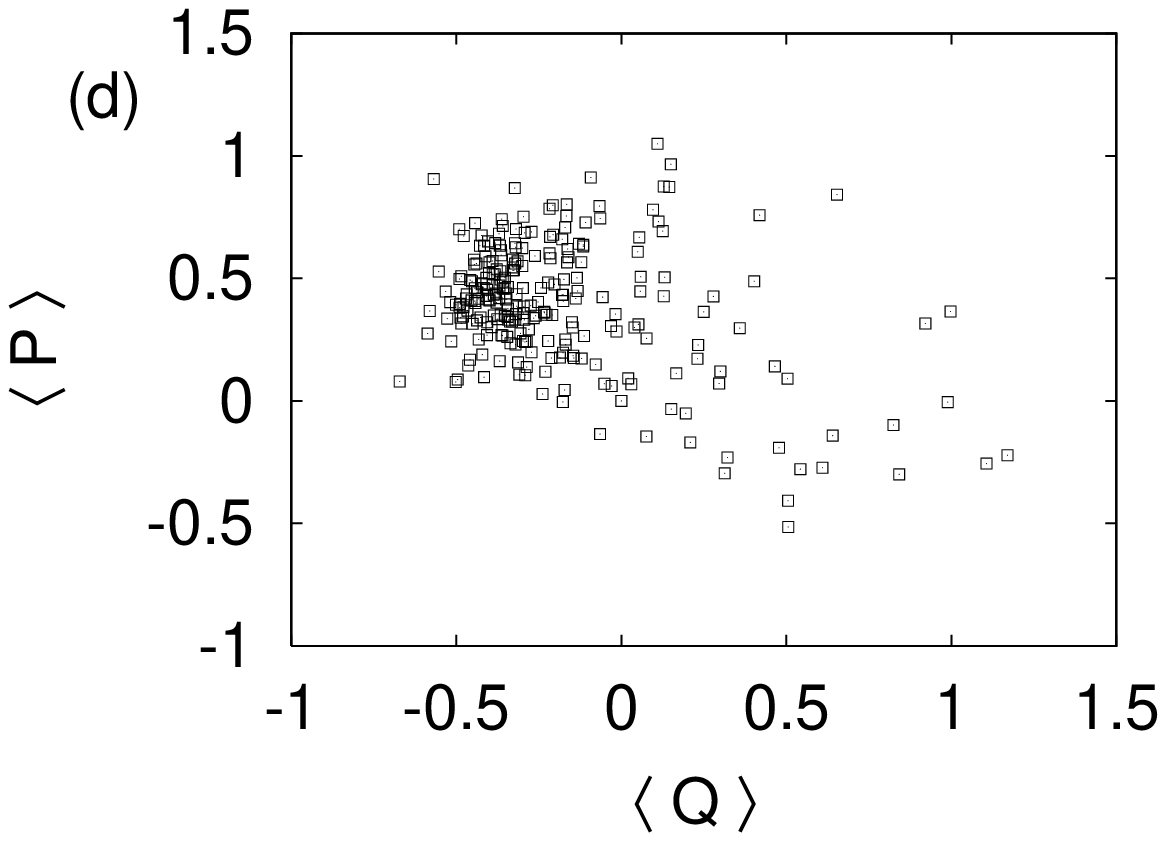}
  \end{tabular}
\caption{\label{fig:1}These are the stroboscopic maps for $(\langle \hat{Q} \rangle,\langle \hat{P} \rangle)$. The each point in these figures represents the data at every
 $2\pi/\Omega$ for a  single realization of complex Wiener process. 
Figures $(a)$, $(b)$, $(c)$ and $(d)$ are for $\beta=0.01$, $0.10$, $0.40$ and $1.00$, respectively.
}
\end{figure*}

First, we show the stroboscopic maps for $(\langle \hat{Q} \rangle , \langle \hat{P} \rangle )$ in the Fig.~\ref{fig:1} for a certain realization of $\zeta (t)$. 
The each point in these figures represents the data at every
$2\pi/\Omega$ for a single realization of $\zeta (t)$. 
The initial state is a pure coherent state $|\alpha =0 \rangle \langle \alpha =0|$, where ${\rm Re} \{ \alpha \} =\sqrt{2} \langle \hat{Q} \rangle$ and ${\rm Im} \{ \alpha \} =\sqrt{2} \langle \hat{P} \rangle$.
These show that a strange attractor appears certainly and the system behaves chaotically in $\beta=0.01$, while it has been lost in $\beta \sim \mathcal{O}(1)$. 
For intermediate case, there remains the remnant of strange attractor. 
We find that the scale of system gets large as $\beta$ goes to zero.
These observations are successful to show the loss of chaotic behavior except for $\beta =0.01$ at least. 
Therefore, let us call that the system is in the classical region for $\beta=0.01$ and in the quantum region for $\beta =1.00$, respectively. 

\begin{figure*}[htbp]
  \centering
  \begin{tabular}{cc}
   \includegraphics[scale=.5]{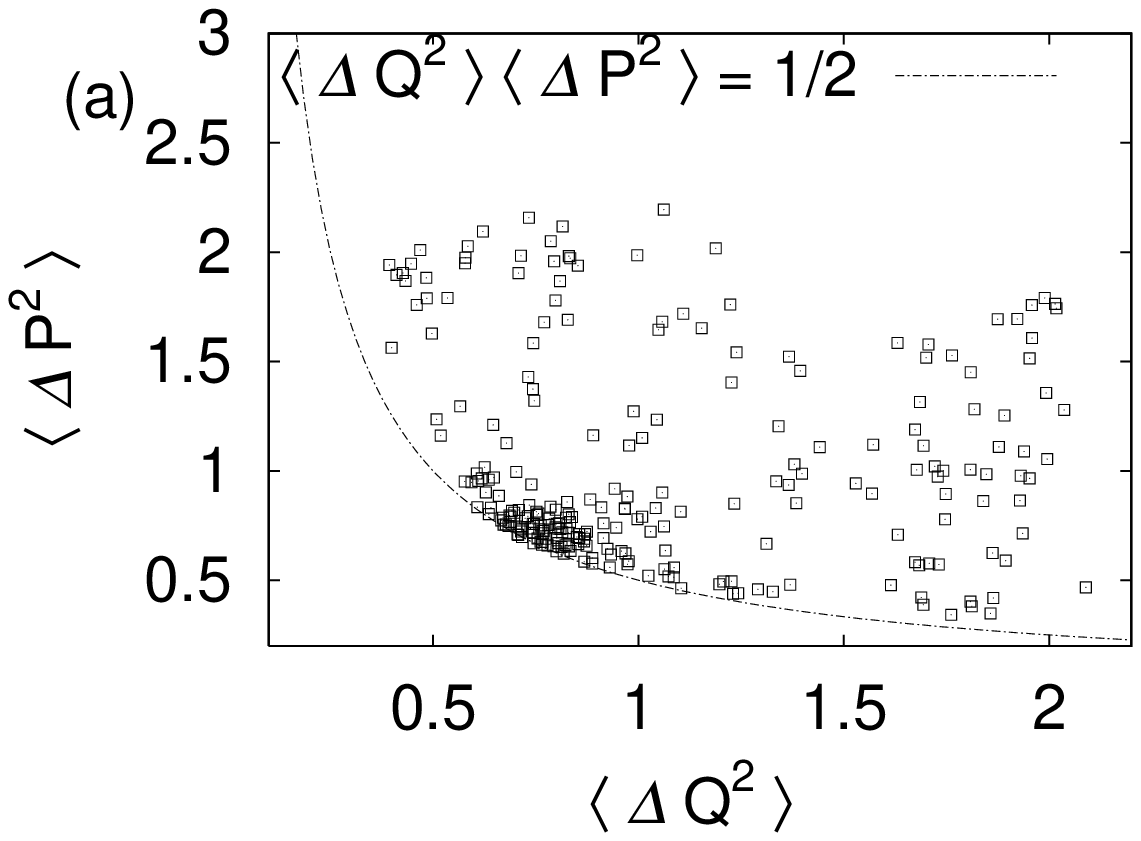} & 
   \includegraphics[scale=.5]{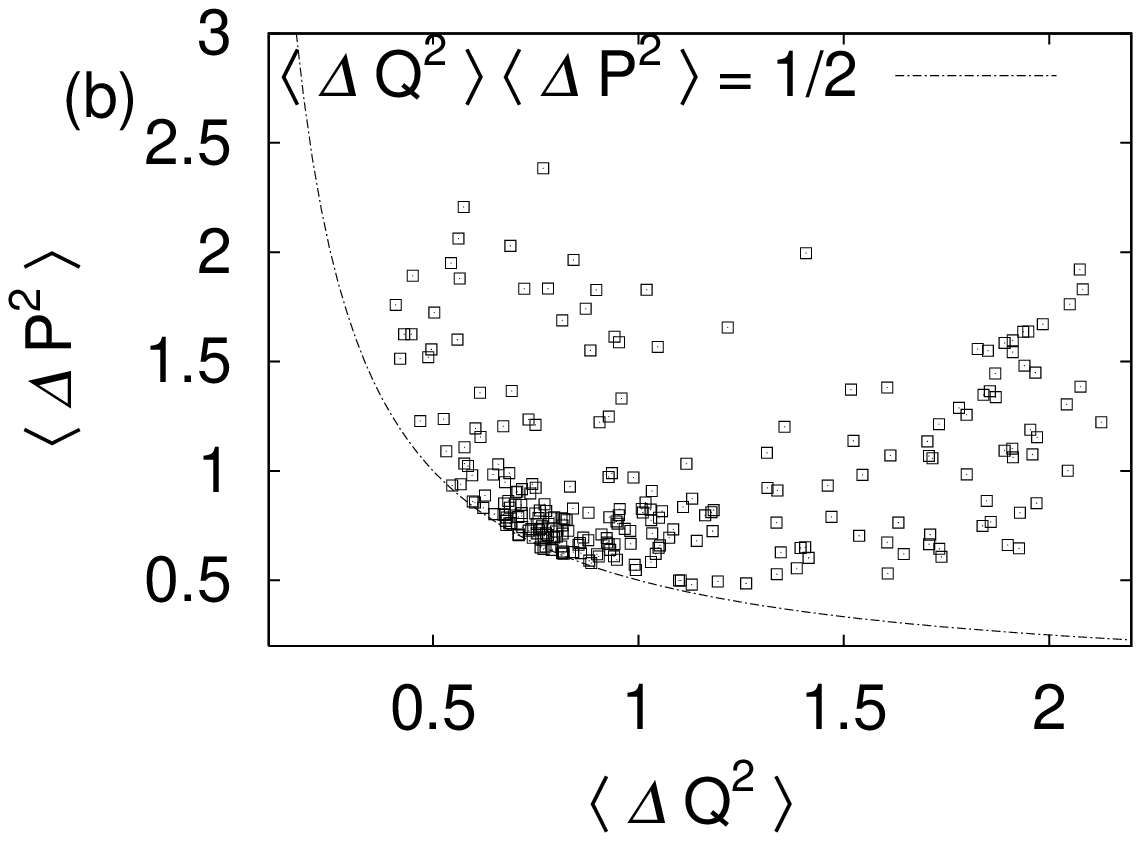}  \\ 
   \includegraphics[scale=.5]{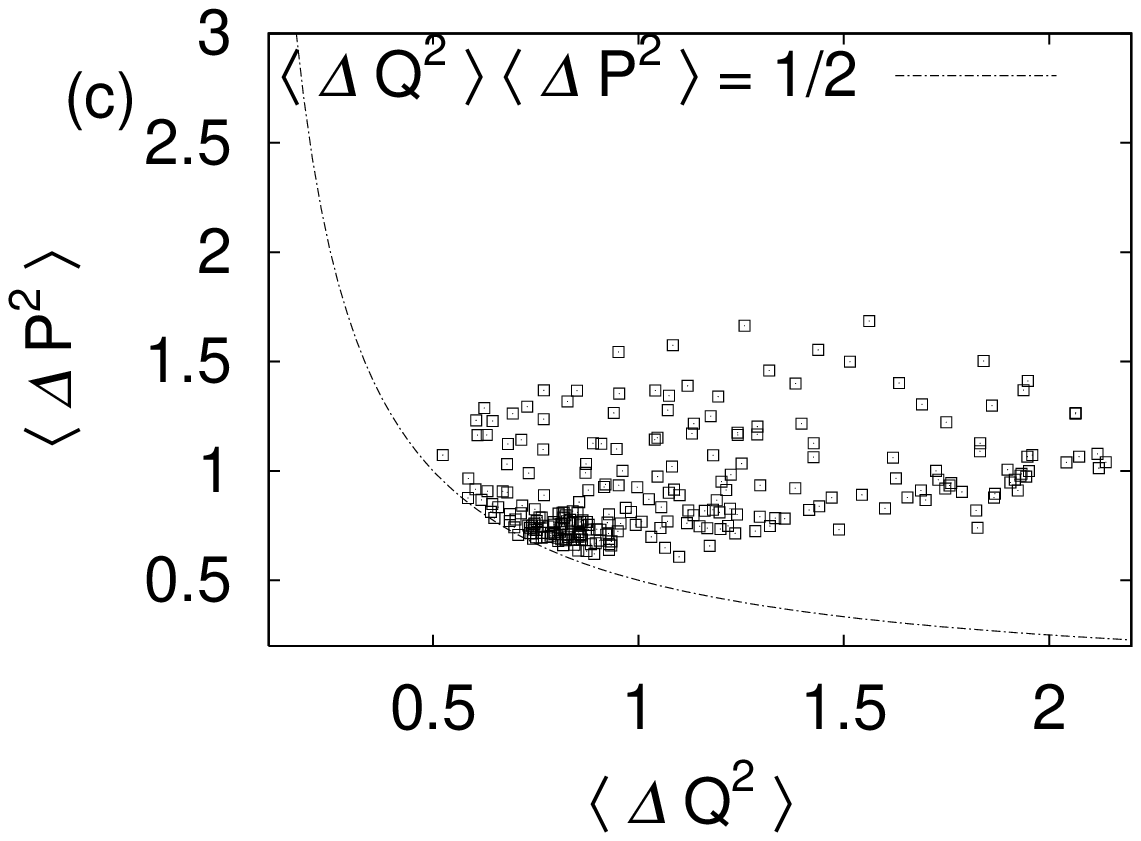}  & 
   \includegraphics[scale=.5]{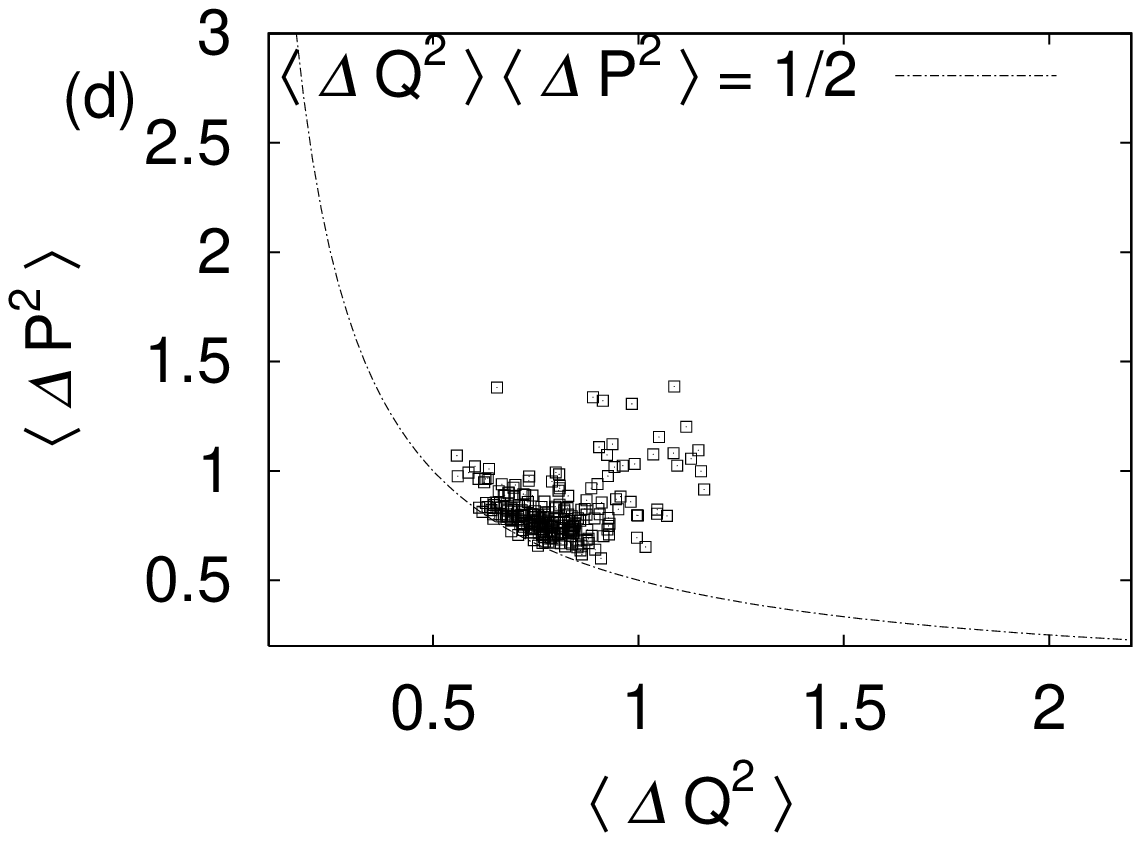}
  \end{tabular}
\caption{\label{fig:2}These are the stroboscopic maps for $(\langle \Delta\hat{Q}^{2} \rangle, \langle \Delta \hat{P}^{2} \rangle)$. The each point in these figures represents the data at every $2\pi/\Omega$ for a single realization of complex Wiener process. Figures $(a)$, $(b)$, $(c)$ and $(d)$ are for $\beta=0.01$, $0.10$, $0.40$ and $1.00$, respectively.
}
\end{figure*}

Next, we show the stroboscopic maps for $(\langle \Delta \hat{Q}^{2} \rangle ,\, \langle \Delta \hat{P}^{2} \rangle )$ in Fig.~\ref{fig:2}, with the same initial condition, $|\alpha=0 \rangle \langle \alpha =0|$.
Since we only investigate the time evolution of {\it stochastic pure state} in the numerical computation, we compute these quantities without averaging over ensemble for $\zeta(t)$.
These results indicate that the state almost preserve minimal uncertainty relation for every $\beta$. 
These results, combined with Fig.~\ref{fig:1}, also verify the argument in Sec.~\ref{Sec:Model} numerically: $\Delta Q/\langle \hat{Q} \rangle$ and $\Delta P/\langle \hat{P} \rangle $ goes to zero if $\beta$ goes to zero.

These analyses without averaging over the ensemble for $\zeta(t)$ have been already studied in Ref.~\cite{Brun}. 
Especially, Fig.~\ref{fig:1} agrees with the results in it.
The results in this section are not first investigation using QSD.
However, these are important to explain our motivation of the analysis
based on `` pseudo-Lyapunov exponent '' in the next section.

\section{\label{Sec:Cross}crossover from classical to quantum behaviors}
In Sec.~\ref{Sec:CR}, we have shown numerically that the quantized Duffing oscillator preserves the characteristic property of the classical dynamics, chaos, in the case of $\beta =0.01$. 
However, this result is inadequate to understand fully the
quantum--classical correspondence in this model due to the following points.
First, it is doubt whether in $\beta=0.01$ the chaotic dynamics 
survives or not, since we only obtain a figure like strange attractor in the stroboscopic maps.
The problem remains even if one claims on the basis of this assertion that the chaotic behavior may occurs in $\beta=0.01$. 
It is not clear at what region of intermediate $\beta=0.01$ and $\beta=1.00$ the classical behavior survives. 
The definition of the classical region or the quantum region is obscure.
Finally, we do not consider the proper quantity related to the behavior of system, as $\beta \to 1.00$.
For example, $\langle \hat{Q} \rangle$ and $\langle \hat{P} \rangle $ in Fig.~\ref{fig:1}
are not the ensemble average $\text{Tr} \{ \hat{Q} \rho \}$ and $\text{Tr} \{ \hat{P} \rho \}$ respectively, are not the values measured in an experiment.

In this section, we introduce a quantity sensitive to initial conditions, ``pseudo-Lyapunov exponent''.
We examine the above three points, based on the analysis of this quantity.
Using this, we can find that there exists the clear crossover from classical to quantum behavior as $\beta \to 1$.

\subsection{\label{subsec:PLE}``Pseudo-Lyapunov exponent''}
We consider the quantity corresponding to instability of classical trajectories which is the most conspicuous characteristic in the classical chaotic systems. 
We define the separation of {\it trajectories} $\Delta (\tau)$ as the following equation: 
\begin{subequations}
\label{Eq:separation}
\begin{equation}
 \Delta (\tau) = \frac{1}{N} \sum_{\{1,2\}} \bigg\{ \delta \overline{Q}_{12}(\tau)^{2} + \delta \overline{P}_{12}(\tau)^{2} \bigg\}^{\frac{1}{2}}, \label{sep_sub1}   
\end{equation}
\begin{eqnarray}
\overline{Q}_{i}(\tau) = \text{Tr}(\hat{Q}\rho_{i}(\tau)),\,\overline{P}_{i}(\tau) = \text{Tr}(\hat{P}\rho_{i}(\tau)),\label{sep_sub2}
\end{eqnarray}
\begin{equation}
 \delta \overline{Q}_{12}(\tau)^{2} =\left( \overline{Q}_{1}(\tau) - \overline{Q}_{2}(\tau) \right)^{2},
\end{equation}
\begin{equation}
 \delta \overline{P}_{12}(\tau)^{2} =\left( \overline{P}_{1}(\tau) - \overline{P}_{2}(\tau) \right)^{2},
\end{equation}
\end{subequations}
where $i=1,2$.
$\rho_{1}(\tau)$ and $\rho_{2}(\tau)$ denote two density matrices for
different initial states $\rho_{1}(0)$ and $\rho_{2}(0)$, respectively. 
Actually, these initial states are the pure coherent states $\lvert \alpha_{i} \rangle \langle \alpha_{i} \rvert,\; (i=1,2)$. 
$\alpha_{i}$ is related to initial condition $(\overline{Q}_{i},\overline{P}_{i})$ by $\alpha_{i} = \sqrt{2} \big( \overline{Q}_{i} + i\overline{P}_{i} \big)$. 
Eq.~(\ref{sep_sub1}) represents the distance in $\text{M} \{ \langle \hat{Q} \rangle \} $--$\text{M} \{ \langle \hat{P} \rangle \}$ plain.
The subscript of $\{1,\,2\}$ in Eq.~(\ref{sep_sub1}) represents the summation over the sets of chosen initial conditions and N is the number of those sets.
The behavior of $\Delta (t)$ is the sensitivity to initial conditions. 
We investigate this behavior in detail, varying $\beta$. 

\subsection{\label{subsec:EPC}Effective Planck cell}
We have to choose a suitable value as the separation $\epsilon \equiv \Delta (\tau=0)$ of two different initial conditions, before the numerical simulations. 
Notice that two points in the phase space are not {\it distinguishable} in the view of quantum mechanics, if they coexist inside the same Planck cell. 
The Planck cell is limited by the Heisenberg's uncertainty relation. 

We explain what is called an the effective Planck cell.
In this model, the commutator $[\hat{Q}, \hat{P}] = [\hat{x}, \hat{p}]/\beta^{2}S = 1$ is fulfilled.
Then the Planck cell has a constant volume of $\Delta Q \Delta P = 1/2$ in the scaled phase space, whereas it has $\Delta x \Delta p = \hbar/2 = \beta^{2} S/2$ in the original phase space. 
With the fixed value of typical action $S$ for the system, the smaller $\beta^{2}$ corresponds to the smaller $\hbar$ and the system exhibits the more classical behavior. 
Thus we can define an effective Planck cell as $\beta^{2}S/2$ ; its
linear size is almost equivalent to $\beta$ in the unit of $\sqrt{S/2}$.

The concept of effective Planck cell gives us how $\epsilon$ should be determined. 
We can consider two kinds of determination of $\epsilon$ for $\beta$: (1) $\epsilon = 0.01$ (fixed), where two points in the phase space are {\it distinguishable} only for the classical region ($\beta = 0.01$). (2) $\epsilon \sim \beta $, where those are {\it distinguishable} for each $\beta$.

\subsection{\label{subsec:CB}Crossover behavior}
In the first, we show that the results of simulation of $\Delta (\tau)$ with $\epsilon$ fixed as $0.01$.
The initial two points are separated at a distance of effective Planck cell for $\beta =0.01$ and coexist inside the cell for the other cases ($\beta \ge 0.1$).
\begin{figure*}[htbp]
  \centering
  \begin{tabular}{cc}
   \includegraphics[scale=.5]{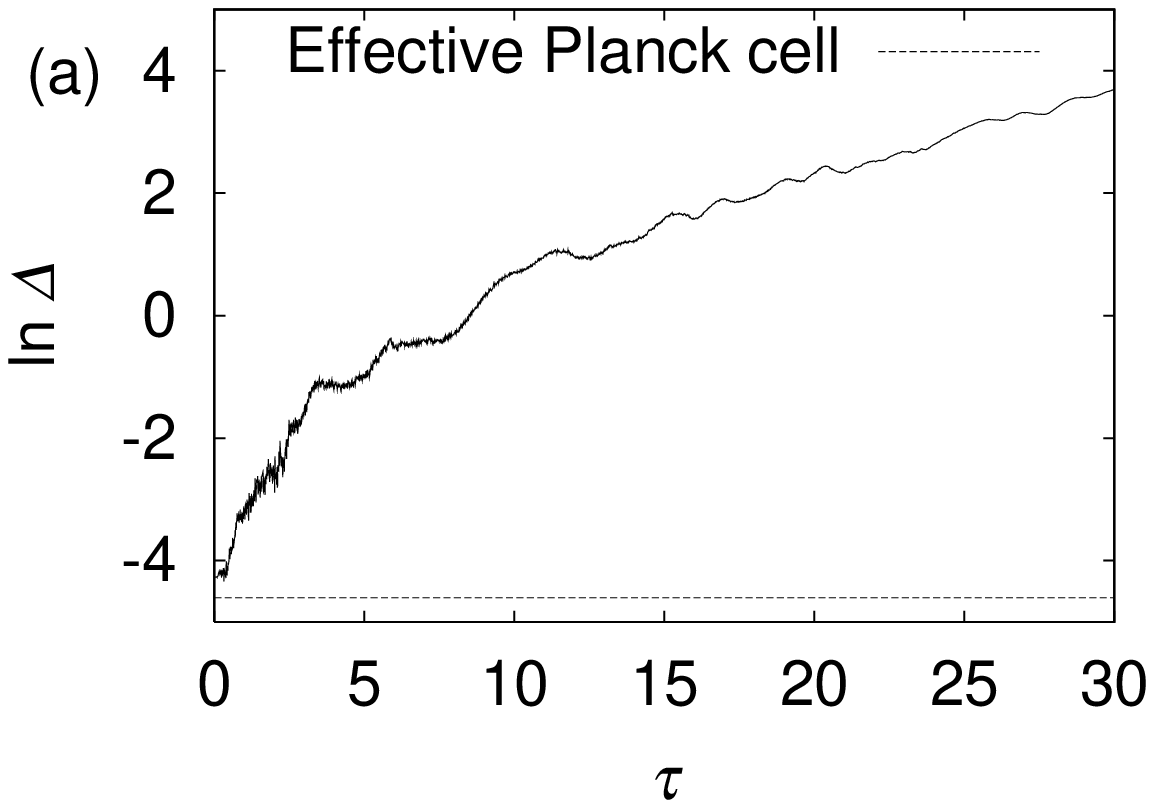} & 
   \includegraphics[scale=.5]{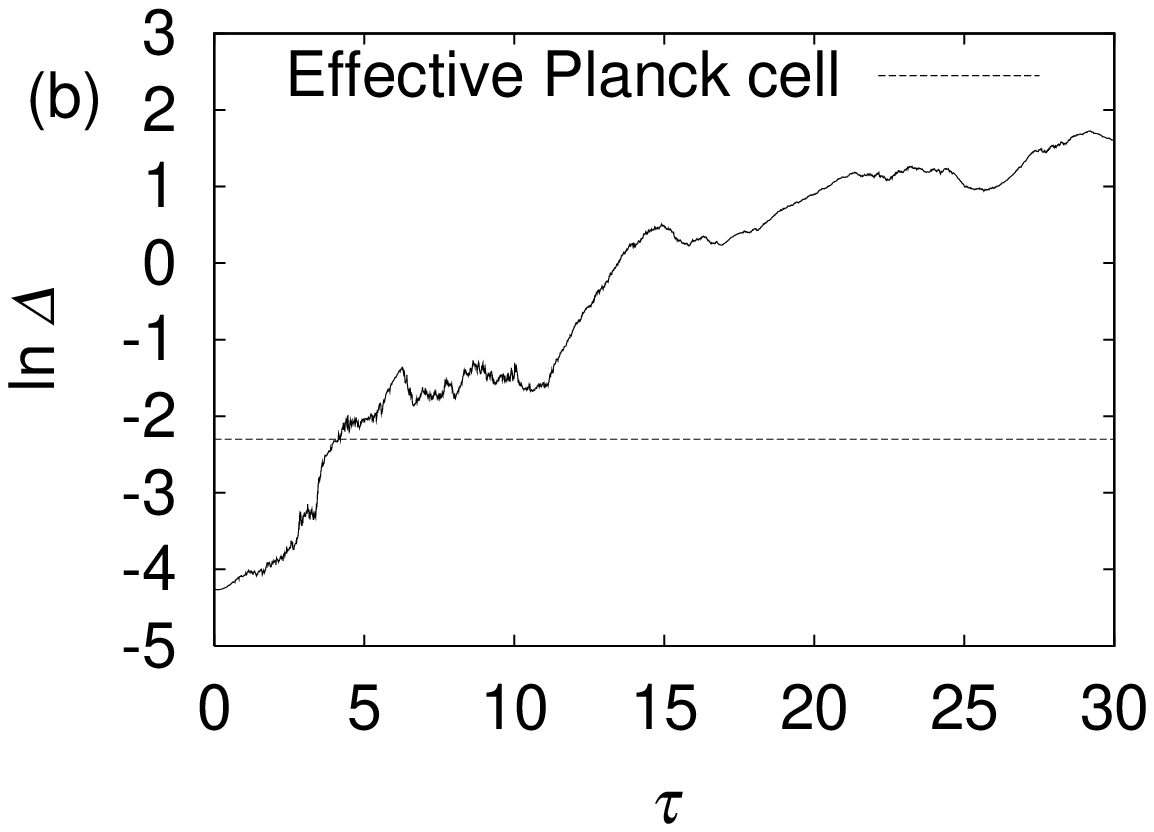}  \\ 
   \includegraphics[scale=.5]{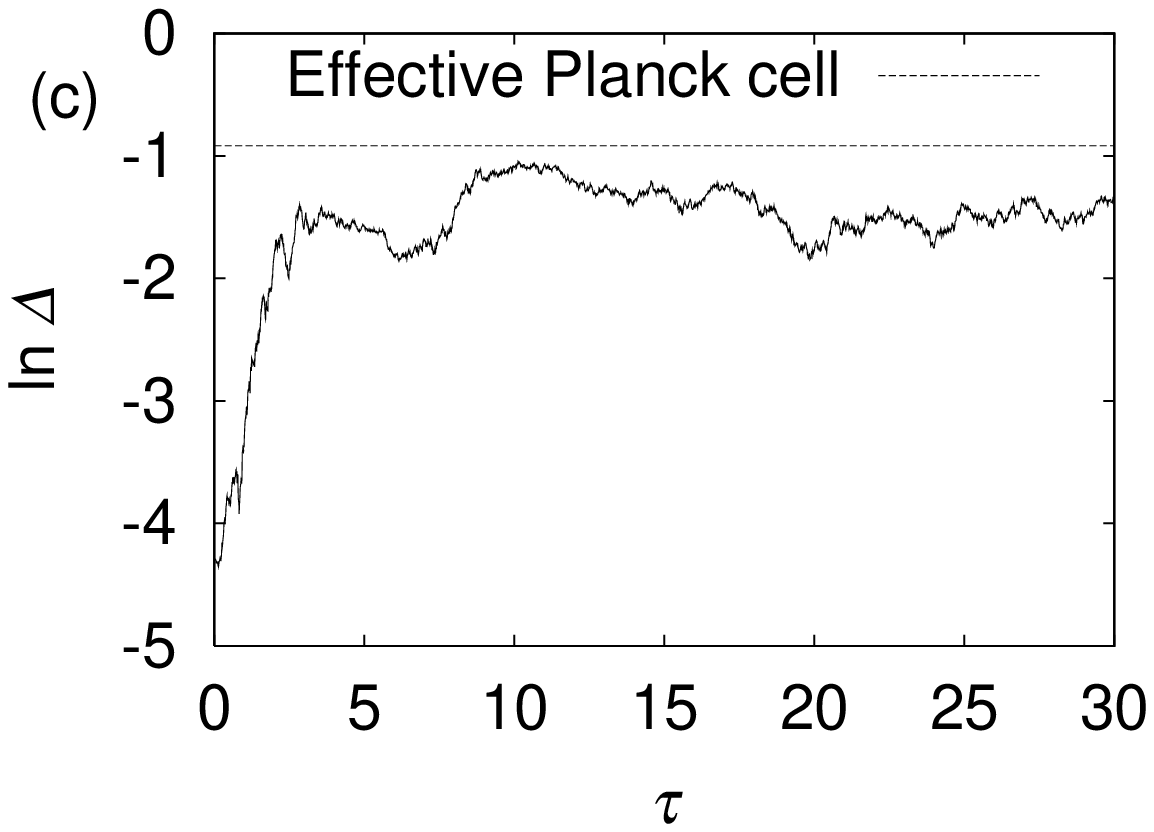}  & 
   \includegraphics[scale=.5]{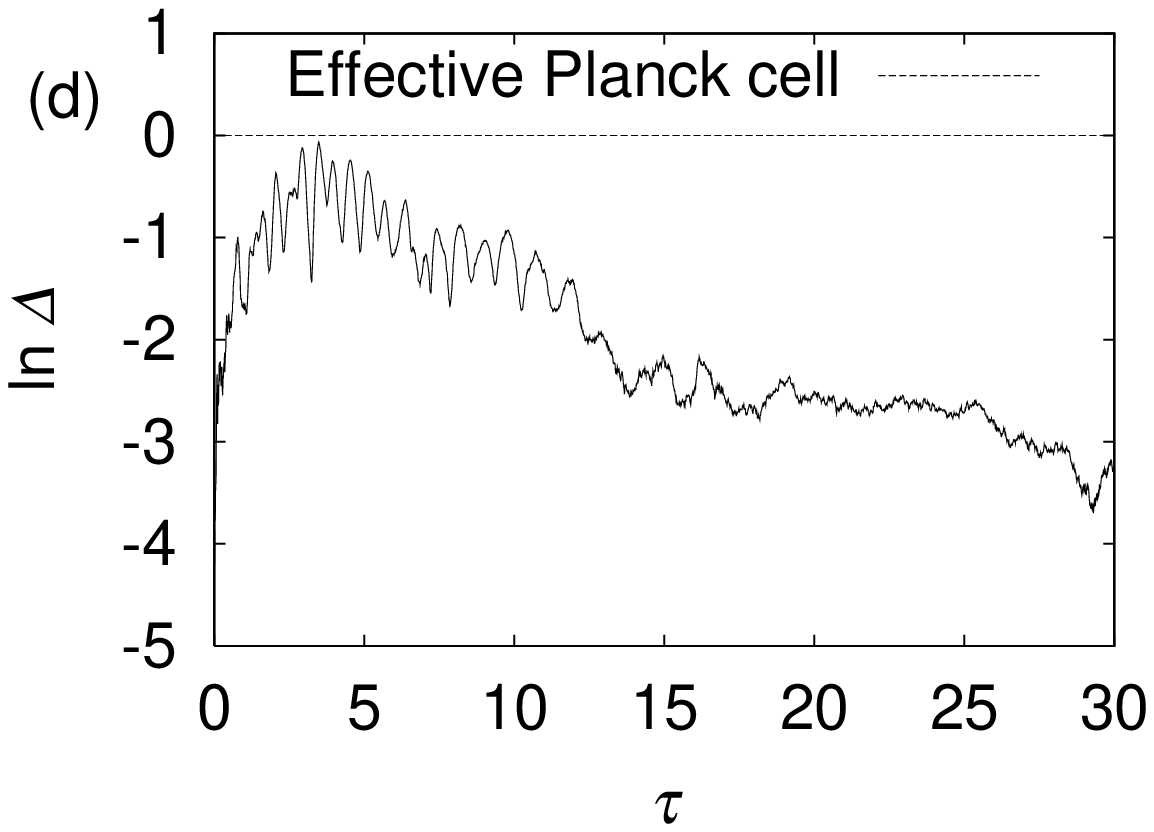}
  \end{tabular}
  \caption{\label{fig:3}
These figures are the time evolution of $\Delta (\tau)$ with $\epsilon$ fixed as 0.01.
Figures (a) and (b) are obtained with single realization of complex Wiener process for each initial condition (20 samples). 
Figures (a), (b), (c) and (d) are for $\beta=0.01$, $0.10$, $0.40$ and $1.00$, respectively.
}
\end{figure*}
In Fig.~\ref{fig:3}(a), we find an exponential increase of $\Delta (\tau)$, a characteristic behavior of chaos. 
This corresponds that maximal Lyapunov exponent is positive in classical mechanics.  
This behavior is also consistent to Fig.~\ref{fig:1}(a). 
These two facts verify that the quantized Duffing oscillator keeps still a chaotic behavior for $\beta=0.01$. 
In Fig.~\ref{fig:3}, we see very different behaviors between (b) and (c)--(d). 
For these values of $\beta$, all  initial two points coexist inside of the effective Planck cell and are {\it indistinguishable} from each other.
Nevertheless, starting from the inside of the effective Planck cell,
$\Delta (\tau)$ for $\beta =0.10$ increases gradually and crosses the size of effective Planck cell after some duration and then increases simply.
This suggests that the remnant of chaotic dynamics still survives for $\beta=0.10$. 
However, $\Delta (\tau)$s for $\beta= 0.40$ and $1.00$ always stay within the effective Planck cell.
The chaotic dynamics has been completely lost in these cases. 
This observation suggests that there exists some critical stage as $\beta $ goes from $0.10$ to $1.00$. 
We argue that there is the crossover from classical to quantum behavior around $\beta \sim 0.40$ due to Figs.~\ref{fig:3} (b)--(d) together with Figs.~\ref{fig:1} (b)--(d).

\begin{figure*}[htbp]
  \centering
  \begin{tabular}{ccc}
   \includegraphics[scale=.5]{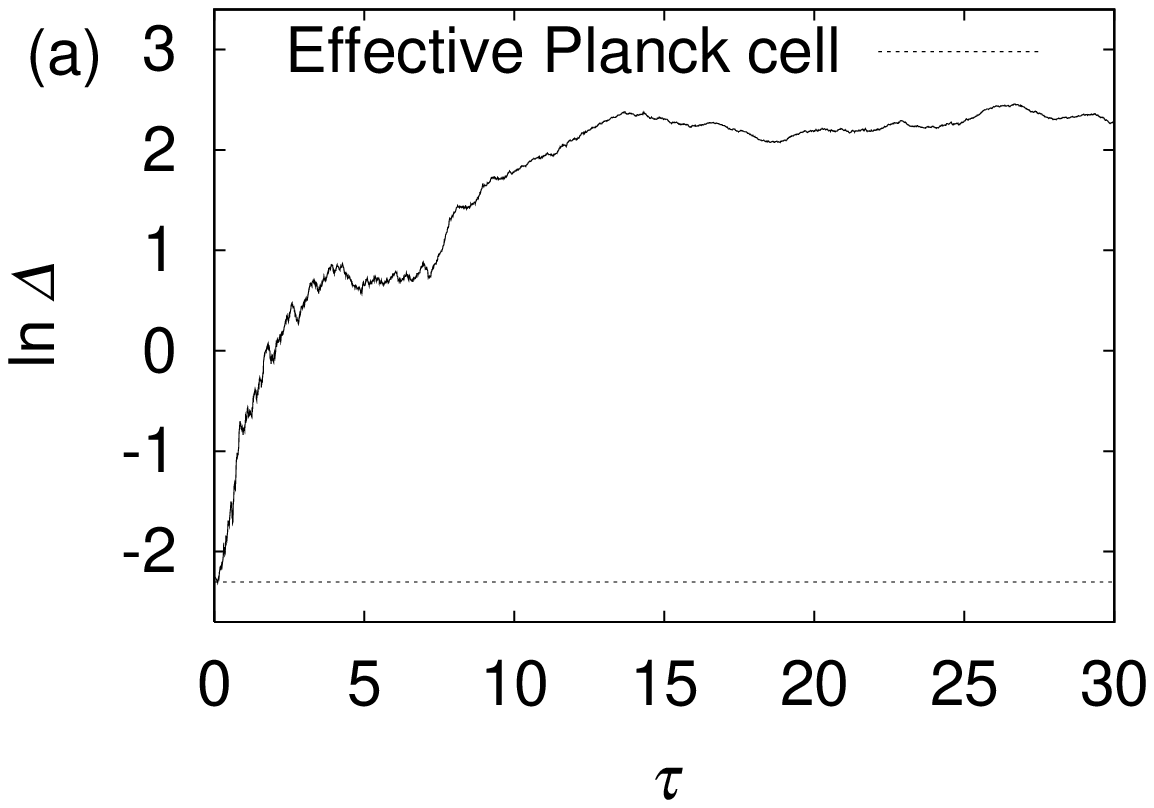} & 
   \includegraphics[scale=.5]{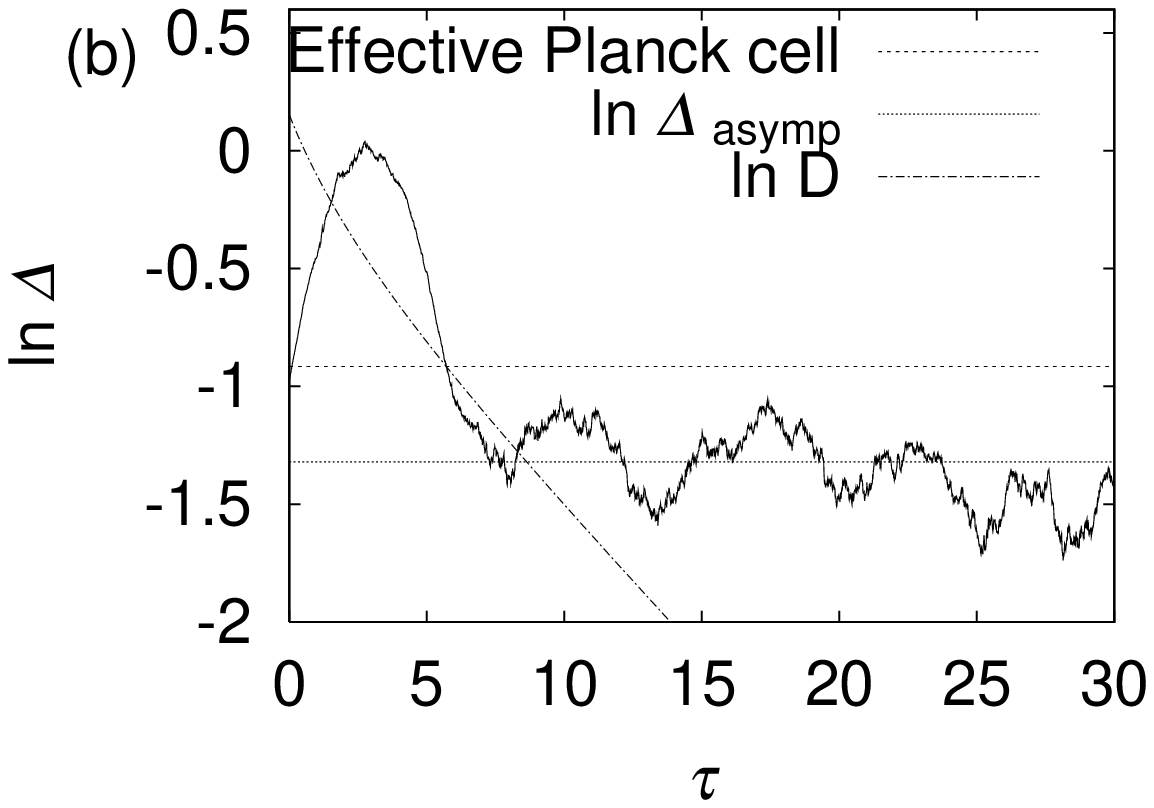} \\ 
   \includegraphics[scale=.5]{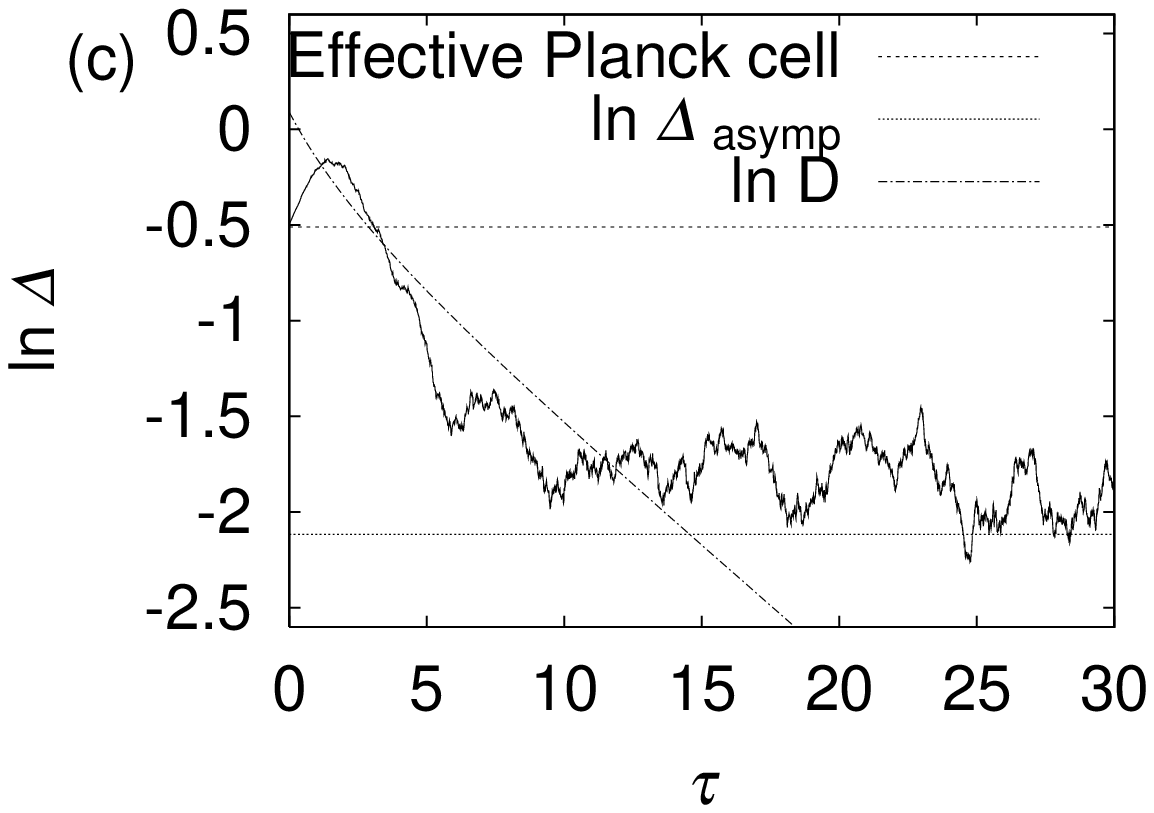} & 
   \includegraphics[scale=.5]{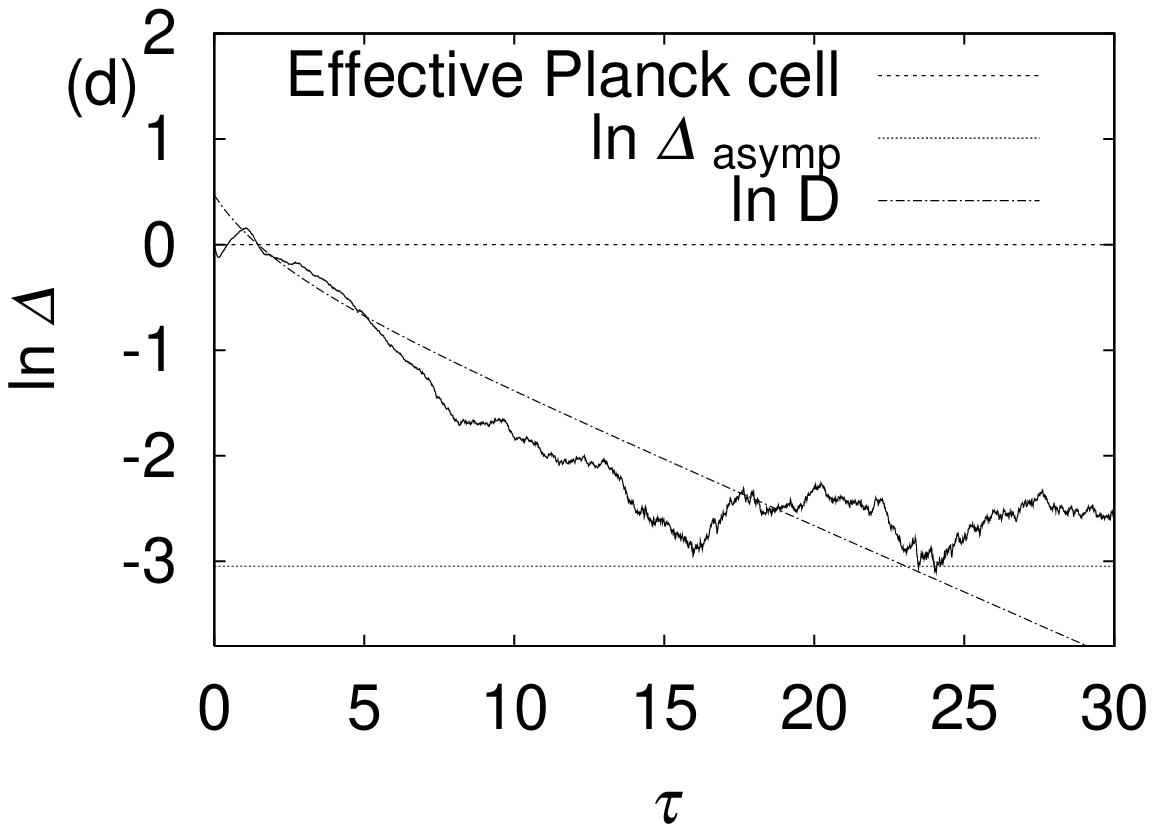} \\
   \includegraphics[scale=.5]{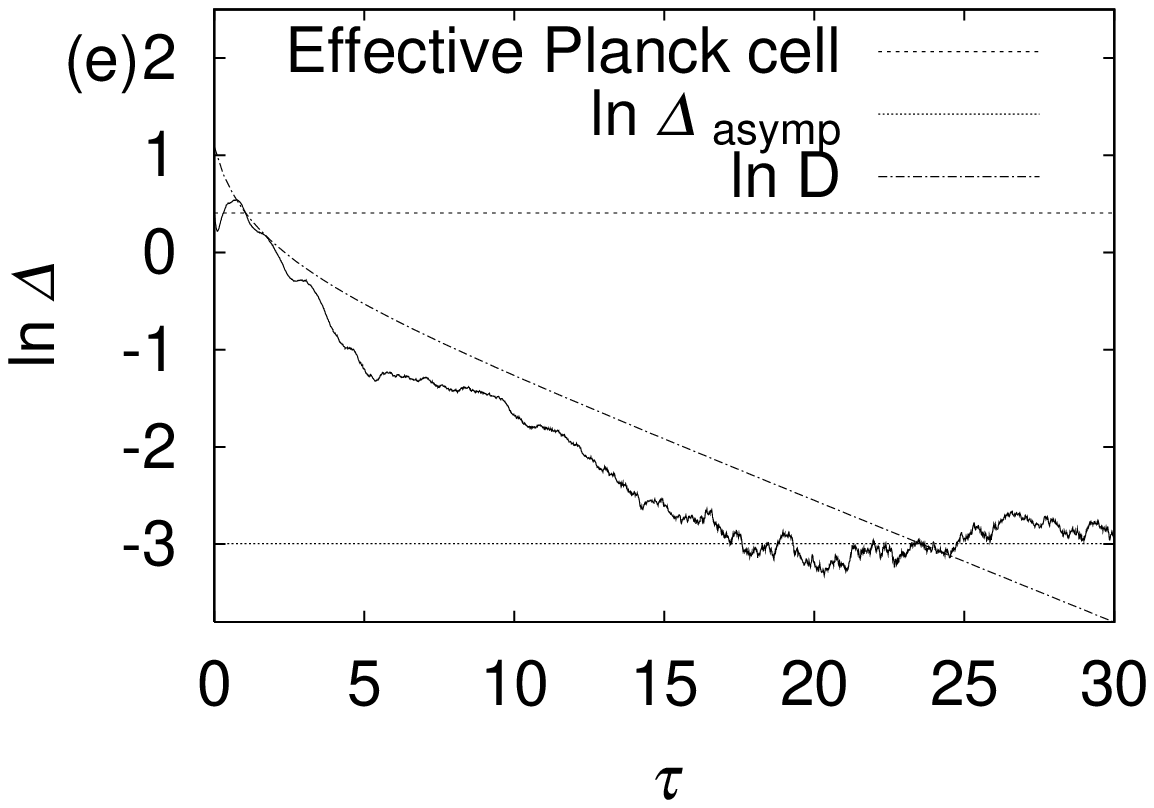} & 	
   \includegraphics[scale=.5]{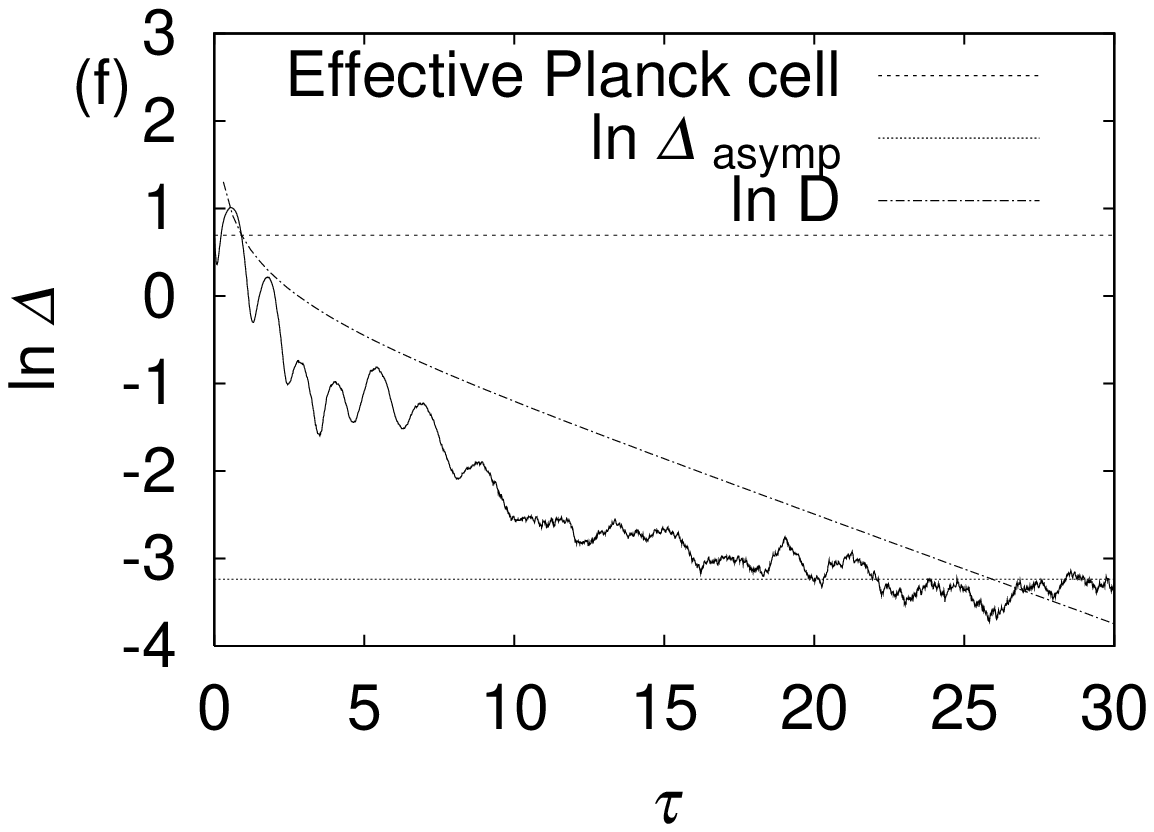}
  \end{tabular}
  \caption{\label{fig:4}
The time evolution of $\Delta (\tau)$ with $\epsilon = \beta$. 
$\Delta_{asymp}$ is the asymptotic value of $\Delta(\tau)$. 
$D$ is the right hand side of Eq.~(\ref{Eq:approD4}). 
Figure (a) is obtained with single realization of complex Wiener process for each initial condition (20 samples). 
Figures (b)--(f) are obtained with averaging over 100 realizations of complex Wiener process for each initial condition (10 samples). 
Figures (a), (b), (c), (d), (e) and (f) are for $\beta=0.10$, $0.40$, $0.60$, $1.00$, $1.50$ and $2.00$, respectively.
} 
\end{figure*}

Let us show other results in the case of $\epsilon \sim \beta$, where
the initial two points are separated at a distance of the effective
Planck cell size.
We compute $\Delta (\tau)$ for $\beta=$ 0.10, 0.40, 0.60, 1.00, 1.50 and 2.00, respectively. 
In Fig.~\ref{fig:4} (a), it is shown that the behavior of $\Delta (\tau)$ for $\beta =0.10$ increases exponentially, which is similar to one in Fig.~\ref{fig:3} (b). 
In Figs.~\ref{fig:4} (b)--(f), it is shown that each $\Delta (\tau)$ takes larger values than the size of effective Planck cell only for a very short period. 
After this period, each $\Delta (\tau)$ always stays within the cell. 
Thus we insist again that there is the crossover behavior at $\beta \sim 0.40$. 

\subsection{\label{subsec:QRDQR}Analyses in quantum and deep quantum regions}
We investigate the case of $\epsilon \sim \beta$ in further detail.
Especially, we focus on the discussion about the case of $\beta \agt 0.40$. 
Hereafter, we call the case of $ \beta >1.00 $ the deep quantum region. 

\begin{figure}[t]
 \includegraphics[scale=.5]{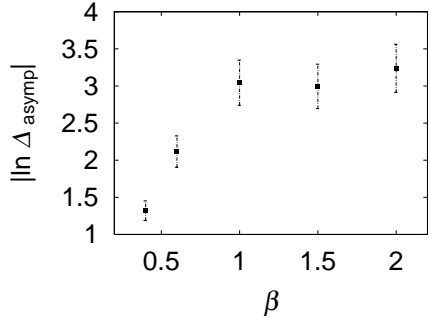}
 \caption{\label{fig:5}The asymptotic value of $\Delta $.}
\end{figure}

In Figs.~\ref{fig:4} (b)--(f), it is found that, except for a very short period after starting time, $\Delta (\tau)$ for each $\beta$ decreases for some duration and tends to approach a certain constant value $\Delta _{asymp}$ asymptotically, which is indicated by the dashed lines. 
In Fig.~\ref{fig:5}, these asymptotic values vs. $\beta$ are shown.
Let us put down $\tau_{asymp}$ and $\tau_{0}$, respectively, when $\Delta
(\tau) $ approaches $\Delta_{asymp}$ and crosses down the boundary of effective Planck cell.

Then, using the QSD, we can obtain an approximate upper bounds of $\Delta (\tau)$ for $\tau \agt \tau_{0}$ as following:
\begin{eqnarray}
\Delta (\tau-\tau_{0}) \alt \left( \left( 1+\frac{1}{\beta^{2}} \right) e^{2\Gamma(\tau-\tau_{0})}-1 \right)^{-\frac{1}{2}}. \label{Eq:approD4}
\end{eqnarray}
In Appendix \ref{app:A}, we derive this estimation in detail. 
Curves calculated by the right hand side of Eq.~(\ref{Eq:approD4}), $D(\tau)$, are represented by broken--dotted lines in Fig.~\ref{fig:4}, which show that Eq.~(\ref{Eq:approD4}) is a good approximation of the upper bound for $\tau_{0} \le \tau \le \tau_{asymp}$. 
In the derivation of Eq.~(\ref{Eq:approD4}), two assumptions are used.
First, we assume that the value of $\Delta (\tau)$ is smaller than the size of effective Planck cell.
Secondly, we assume that the dissipative effect dominate over the systematic time evolution by the Hamiltonian for some time duration  from the starting time. 
Thus, $\Delta (\tau)$ for $\tau_{0} \le \tau \le \tau_{asymp}$ is
probably described mainly by the dissipative effect. 
Eq.~(\ref{Eq:approD4}) should not work well for $\tau > \tau_{asymp}$, where $H=0$ is not a good approximation and there exists many other sources confining $\Delta (\tau)$ within a small value.
$\Delta (\tau)$ may become the constant value, $\Delta_{asymp}$, related to the inherent property of system.
In classical mechanics, the dissipative chaotic systems like Duffing oscillator generate the chaotic dynamics due to the coexistence of dissipative effect and periodic external force.
In the quantized Duffing oscillator, if the action of system is much greater than $\hbar$, it seems similar to classical, i.e., the existence of dissipation is very important for occurrence of chaotic dynamics. 
On the other side, if the action of system is smaller than $\hbar$, that is, in quantum and deep quantum case, the analysis of $\Delta_{asymp}$ and $\tau_{asymp}$ suggests that the effect of dissipation suppresses even the occurrence of chaotic behavior. 

In Fig.~\ref{fig:5}, it is found that $\Delta _{asymp}$ takes a certain constant value not depending on $\beta$ in the deep quantum region. 
According to the above discussion, we find that $\Delta (\tau)$ well characterizes the behavior of system between the classical and quantum regions, but not in the deep quantum region. 
It will be necessary to investigate other quantities, for example, the higher moments and Wigner function, in order to analyze the behavior in the deep quantum region in detail. 
However, $\Delta (\tau)$ is an effective quantity enough to investigate the crossover behavior between the classical and quantum regions.

Finally, we explain our method estimating $\tau_{asymp}$. 
First, we determine $\Delta_{asymp}$ and $\tau_{0}$ for each $\beta$, using the results of simulation.
Secondly, we estimate $\tau_{asymp}$ based on the relation $D(\tau_{asymp}-\tau_{0}) = \Delta_{asymp}$ for each $\beta$. 
The results are $(\beta,\, \tau_{0},\, \tau_{asymp})$ = $(0.40,\, 5.71,\, 8.63)$, $(0.60,\, 2.88,\, 11.9)$, $(1.00,\, 1.45,\, 19.4)$, $(1.50,\, 1.04,\, 23.5)$ and  $(2.00,\, 0.91,\, 25.9)$. 
It is found that $\tau_{asymp}$ takes a constant value not dependent on $\beta$ in the deep quantum region. 
This result is consistent with an approximation of $D$ at large $\tau$:
$$
D(\tau - \tau_{0}) \sim \Big\{ 1-\mathcal{O} \Big(\frac{1}{\beta^{2}}
\Big) \Big\} e^{-\Gamma (\tau - \tau_{0})}.
$$
\begin{figure}[t]
 \includegraphics[scale=.5]{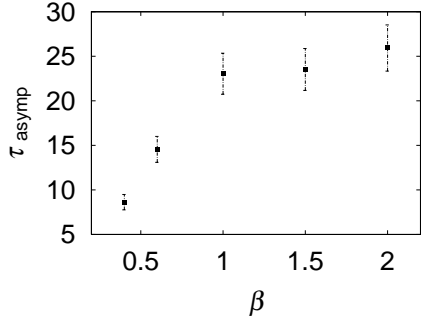}
 \caption{\label{fig:6}The asymptotic time $\tau_{asymp}$.}
\end{figure}

\section{\label{sec:Summ}Summary}
We have shown several numerical results for quantized Duffing
oscillator, and discussed the crossover from classical to quantum
behavior, based on the sensitivity to initial conditions, ``
pseudo-Lyapunov exponent ''. 
In our discussion, it is important how $\Delta (\tau)$ behaves as the scaling parameter $\beta$ varies. 
We have roughly defined $\beta =0.01$ as the classical region, $\beta=1.00$ as the quantum region and $\beta >1.00$ as the deep quantum region. 
We have found the following points, analyzing the numerical results. 
In the classical region, we prove that the chaotic behavior appears certainly, since $\Delta (\tau)$ increases exponentially. 
In the quantum region, we show that it has been lost completely, since $\Delta (\tau)$ does not increases exponentially and takes a certain constant value asymptotically. 
Notice that there is a clear crossover behavior as $\beta$ increases from 0.01 to 1.00; $\Delta (\tau)$ for $\beta =0.10$ still increases exponentially, but such a behavior has been lost around $\beta =0.40$.  
For intermediate $\beta$ we can insist that the system is classical even in the case of $\beta =0.10$. 
In the deep quantum region, $\Delta (\tau)$ is not a suitable quantity to characterize the behavior of system. 
It is probably necessary to investigate other quantities in order to analyze the behavior in the deep quantum region. 
However, $\Delta (\tau)$ is an important quantity enough to investigate the crossover behavior between the classical and quantum regions, since it clarifies the nontrivial crossover behavior. 
Moreover, we have understood why the chaotic behavior has been lost as $\beta \to 1.00$ in the view of dissipative quantum systems; the effect of dissipation suppresses the occurrence of chaotic behavior in the quantum region. 
The effect of Hamiltonian occur after $\tau_{asymp}$. 

The problem of dissipative quantum chaos has many topics of the
foundation for quantum
theory\cite{Adamyan,Bhattacharya,Braun,Haake,Miller,Habib,ZurekPaz,Zurek}.
Indeed, there are still many problems for the quantum-classical correspondence.
For example, we have to estimate quantitatively how much classical trajectory exerts an effect on the quantum system.
$\Delta (\tau)$ takes a meaning value in the classical region, while it is not clear what significant the value of $\Delta(\tau)$ has in the quantum region.
Nevertheless, this analysis based on ``pseudo-Lyapunov exponent'' clarifies the crossover in the quantized Duffing oscillator at first. 
This point is quite different from the previous work\cite{Brun}.

\section*{Acknowledgment}
The authors would like to thank T. Brun and R. Schack who offered us the algorithm of QSD used in this work. 
This work is supported partially by the Grant-in-Aid for COE Research and that for Priority Area B (\#763), MEXT. 
The authors thank the Yukawa Institute for Theoretical Physics at Kyoto University. 
Discussions during the YITP workshop YITP-W-02-13 on ``Quantum chaos: Present status of theory and experiment'' were useful to complete this work. 

\appendix
\section{\label{app:A}The derivation Equation (\ref{Eq:approD4})}
We show the derivation of Eq.~(\ref{Eq:approD4}).
To work out this, we should notice that $\Delta (\tau)$ is smaller than the size of effective Planck cell for $\tau > \tau_{0}$. 
Therefore, it is suggested that two wave packets starting from two different initial conditions locate very closely and almost overlap respectively. 
Thus $\Delta (\tau)$ for $\tau > \tau_{0}$ can be approximated by $D(\overline{Q},\, \overline{P},\, \tau)$ characterized by the spread of wave packet for only one initial condition in the phase space. 
Of course, this approximation is valid for $\tau > \tau_{0}$.

We define $\{ D(\overline{Q},\, \overline{P},\, \tau)\}^{2}$ as
$\text{M} \{ \sigma(\hat{a}^{\dag} ,\hat{a}) \} \equiv \text{M} \{
\langle \hat{a}^{\dag} \hat{a} \rangle - \langle \hat{a}^{\dag} \rangle
\langle \hat{a} \rangle \}$ with $\hat{a}= \left( \hat{Q} +i \hat{P} \right)/\sqrt{2}$.
In \cite{Percival}, $\text{M} \{ \sigma(\hat{a}^{\dag} ,\hat{a}) \}$ is introduced in order to investigate the localization in framework of QSD.
Notice that this quantity is different from $\text{Tr} \{ \hat{a}^{\dag} \hat{a} \rho \}-\text{Tr} \{ \hat{a}^{\dag} \rho \} \text{Tr} \{ \hat{a} \rho \}$. 
$\text{M} \{ \sigma(\hat{a}^{\dag} ,\hat{a}) \}$ can be only calculated by means of QSD.
Considering the average over the set of initial conditions, we can write 
\begin{eqnarray}
\Delta (\tau) \approx \int d\overline{Q} d\overline{P} ~\mu (\overline{Q},\, \overline{P}) D(\overline{Q},\, \overline{P},\, \tau). \label{Eq:approD1}
\end{eqnarray} 
$\mu (\overline{Q},\, \overline{P})$ is the distribution function of initial condition. 
Since the classical orbits move over some finite region in the phase space, e.g., strange attractor, the integration is limited by some finite volume $V$.
If $\mu(\overline{Q},\,\overline{P})$ is uniform distribution, 
\begin{eqnarray}
\Delta (\tau) \approx \frac{1}{V}\int d\overline{Q} d\overline{P} ~D(\overline{Q},\,\overline{P},\, \tau). \label{Eq:approD2}
\end{eqnarray}

If Hamiltonian $\hat{H}=0$, we can derive the following equation: 
\begin{eqnarray*}
\frac{d}{d\tau} \text{M} \{ \sigma(\hat{a}^{\dag} ,\hat{a}) \}&=& - 2\Gamma \text{M} \{ \sigma(\hat{a}^{\dag} ,\hat{a}) \} - 2\Gamma \text{M} \{ \sigma(\hat{a}^{\dag} ,\hat{a})^{2} \} \\& & -2\Gamma \text{M} \{ \sigma(\hat{a}^{\dag} ,\hat{a}^{\dag})\sigma(\hat{a},\hat{a}) \} \\ &\le&-2\Gamma \text{M} \{ \sigma(\hat{a}^{\dag} ,\hat{a}) \} -2\Gamma (\text{M} \{ \sigma(\hat{a}^{\dag} ,\hat{a})\} )^{2}. 
\end{eqnarray*}
In the case of equality in the above equations, this relation is Ricaci's differential equation.
Put $u \equiv \text{M} \{ \sigma(\hat{a}^{\dag} ,\hat{a}) \}$, and  notice that u is positive for all $\tau$. 
Thus, 
\begin{eqnarray*}
& &\frac{d}{d\tau} \left( \frac{1}{u} \right) \ge2\Gamma \left( 1+\frac{1}{u} \right) \\ &\Longleftrightarrow& \frac{1}{u(\tau)} - \frac{1}{u(\tau_{0})}\ge 2\Gamma \int_{\tau_{0}}^{\tau} ds \left( 1+\frac{1}{u(s)} \right) \\ &\Longleftrightarrow&  \frac{1}{u(\tau)} \ge \left(1+\frac{1}{u(\tau_{0})} \right) e^{2\Gamma (\tau-\tau_{0})} -1 \\ &\Longleftrightarrow& u(\tau) \le \frac{1}{Ce^{2\Gamma (\tau -\tau_{0})}-1},
\end{eqnarray*}
where $C=1+1/u(\tau_{0})$. 
Therefore we obtain the upper bound of $\Delta (\tau)$ for for $\tau >\tau_{0}$ in the following equations:
\begin{eqnarray}
\Delta (\tau) &\alt& \frac{1}{V} \int d\overline{Q} d\overline{P}~\left( \left(1+\frac{1}{\beta^{2}} \right) e^{2\Gamma(\tau-\tau_{0})}-1 \right)^{-\frac{1}{2}} \nonumber \\ &=& \left( \left( 1+\frac{1}{\beta^{2}} \right) e^{2\Gamma(\tau-\tau_{0})}-1 \right)^{-\frac{1}{2}}, \label{Eq:appresult}
\end{eqnarray}
where we use $\Delta (\tau_{0}) = \beta$. 
Eq.~(\ref{Eq:appresult}) is just Eq.~(\ref{Eq:approD4}).

\bibliography{crossover}

\end{document}